# Time Efficient Data Migration among Clouds

Syeda Munazza Marium, Liaquat Ali Thebo, Syed Naveed Ahmed jaffari
Computer System Engineering Department
Mehran University of Engineering & Technology
Sindh Pakistan
munazza.syed_1@yahoo.com,
liaquat.thebo@faculty.muet.edu.pk,
Naveed.jaffari@faculty.muet.edu.pk

Muhammad Hunain Memon
School of Information Science and Technology
University of Science and Technology of China,
Hefei, China
hunainmemon@ieee.org

*Abstract*— **Cloud computing is one of the chief requirements of modern IT trade. Today's cloud industry progressively dependent on it, which lead mutually abundant solutions and challenges. Among the numerous challenges of Cloud computing, cloud migration is one of the major concern, and it is necessity to design optimize solutions to advance it with time. Data migration researchers attempt to move data concerning varying geographical locations, which contain huge data volumes, compact time limit and problematical architectures. Researchers aim to transfer data with minimal transmission cost and used various efficient scheduling methods and other techniques to achieve this objective. In former research struggles, numerous solutions have proposed. In our proposed work, we have explore the contextual factors to accomplish shorter transmission time. Entity Framework Core technology is utilize for conceptual modelling, mapping and storage modelling. Meant for minimum transmission cost Object Related Mapping is designated. Desired objective to achieve time efficiency during data migration has been accomplished. Results obtained when data transmission occur among azure and gearhost cloud with implementation of proposed framework with some size limitations.**

## I. Introduction

This Cloud computing emerging very speedily as a pervasive computational paradigm. It is a software package that continuously gaining recognition and acceptance as a resource of economical and dependable computing solution and services through internet. According to the statistical analysis worth of cloud computing market is billion dollars. It has three models which are IaaS, PaaS, SaaS (Infrastructure as a service), (platform as a service) and (software as a service) respectively. Whereas Iaas provide (storage, network, CPU, etc.) as a service.

Nowadays, a large number of applications store their data on storage servers, these applications include sensor networks, search engine cluster, video on demand servers and grid computing. In such application data migration among clouds demands lots of services, because every server run on different protocol.

In this paper they focused on an existing cloud computing technology and also forthcoming inquiries in this area and in different cloud environment, explore, service equivalence [35].This researched focused on an issue called cloud migration. Transferring data from one cloud to another with efficiency and operational processing is aimed here. Proposed online lazy migration (OLM) algorithm and a randomized fixed horizon control (RFHC) algorithm as a solution for cost-minimization problem in data migration [31]. This work is dedicated to key challenges emerged when dealing with Iaas Infrastructure as a service and networking architecture of cloud like Software-defined networking (SDN) and other architectures [29]. In this paper author investigates mobile cloud architecture and present critical analysis over application model classification, decision making entities, execution delay, cloud application models and mobile synchronization policies [32]. Author addresses the security issues when we upload data on cloud and migrate it (i-e privacy-preservability, accountability, Integrity, confidentiality, availability) [30]. This paper covers energy efficiency domain of cloud computing separated into two domains Server and network. It determine and show correlation among various domains of ICT related to energy efficiency [24]. Here author puts light on the cloud interoperability matter, discuss band of challenges like resource availability and scalability, avoiding vendor lock, interoperability, low latency and other legal issues [27]. This paper is a review of Cloud Computing (CC) and Information Technology Outsourcing (ITO) address variance among infrastructure and software services, utilization of cloud self-service and emerging role of IT in it and coded contributing elements which effect these decision [23]. Following paper author discourse about an important issue, the heterogeneous mobile platform for cloud computing. Performed examination on origins of Mobile Cloud Computing (MCC) heterogeneity factors like vendors, platform, network API and other features and recognized many challenges, to overcome these limitations analyses different architectures SOA, virtualization and middleware, etc. [25]. Author considers both spectrum efficiency and pricing efficiency of cloud and analyse power and interference management in cloud network. Proposed iterative algorithm as a solution to achieve steadiness [21]. This survey focused resource scheduling architecture of cloud computing. The survey classified resource scheduling in three categories (a) application layer scheduling (b) virtualization layer scheduling (c) deployment layer scheduling [19]. Increasing traffic requirements decrease energy efficiency of a cloud. To increase energy efficiency three approaches are presented (a) MIMO (b) Dynamic spectrum access technology (c) Design frequency reuse scenario by creating smaller cells. Cloud Radio Access Network (C-RAN) with multimode support is a new model to achieve efficiency.[17]Computation offloading classify into three manners .(a) remote cloud service (b) opportunistic ad hoc cloud service and (c) connected ad hoc cloud service [18]. Virtual migration scenario is demonstrated for private cloud of organization whose applications reside on network .VM produce efficiency by removing data duplication during data migration in active and in active state [15]. Cloud resource management and scheduling performance achieve by implementing QoS-constrained algorithm on static and dynamic load which improve resource utilization and security [16]. Analyseproposed big data migration framework in current scenario using APACHE, HADOOP and SPARK utilizing various data processing schemes and machine learning algorithms, and present 5G wireless architecture prototype to process huge amount of data[9] .Flexibility of CC achieve by infrastructure virtualization. Mainly it depends on cloud computing infrastructure and design and development of application [13].

## II. CLOUD COMPUTING

Earlier Network Diagrams used cloud as a symbol to represent Wide Area Network (WAN) with this context word cloud used with the internet. Now term cloud and computing used together, although cloud represent internet and computing embody services .It is blend of several amenities, which involve application development platform, shared pool resources, system management, Scalability, and multiple other services enlisted in Figure 1.

### A. Key Advantages of Cloud Computing

Cloud computing delivers computational services like software integration, server space, database, storage space, backup, recovery, analytics and etc. These services are paid according to the demand and usage.

- Budget Effective: CC is cost proficient. There is no need to buy costly hardware and software. You can use them on cloud, according to Pay-as-you-go Plans. It saves money and provide well-organized resource utilization as per need.

- Universal scale: Cloud data centers are universal they are spread globally so accurate amount of bandwidth, resources can be delivered at any time. From any geographic area, one can access cloud resources.

- Throughput: Maximum output can be achieved because IT team has no burden to manage hardware, software integrity and resource handling. So entire time can be spent to achieve business goals.

- Speed: It is so quick to get access of any hardware and software service according to demand on cloud in just few minutes. So it makes so flexible to acquire such resources.

- Performance: The cloud data centers equipped with the latest technology hardware and software resources and always up to date. So these resources deliver high performance computing for your business goals.

- Trustworthiness: CC comes with backup, data recovery and disaster plan which introduce more reliability and less risk factor in computing when we are using resources from the cloud. Cloud provider network mirrored our data in a secure environment.

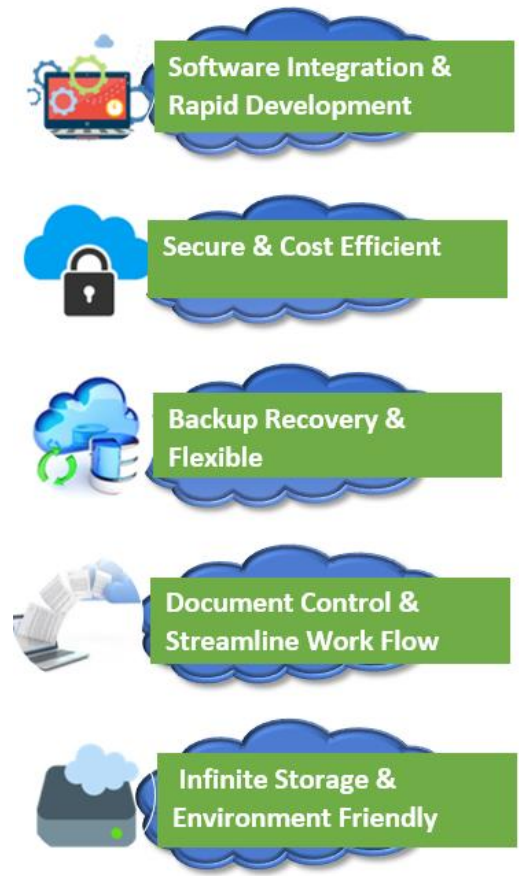

Figure 1. Cloud Computing Advantages.

### B. Categories of Cloud Services

CC services mainly divided into three classes. Every so often is known as computing Stack because they are in stack manner. They are dependent, but different from each other explained below and shown in Figure 2.

- Infrastructure-as-a-service (IaaS): Provide virtualized computing resources like network bandwidth, processor cycles and host virtualized infrastructure using hypervisor. This class can rent infrastructure, network, server, storage, virtual machines etc. from a cloud as according to your demand. IaaS vendors include Amazon, Rackspace, Cloud Foundry.

- Platform as a service (PaaS): It deliver adaptable and optimal environment. For development purpose this service is used to develop applications on cloud platforms so there is no need to worry about databases, storage, network and other resources essential for software development some examples of PaaS vendors include Microsoft Azure, Amazon, Force.com.

- Software as a service (SaaS): This distribution model is persistent, accessible, and scalable eliminate expense of Licensing, maintenance provisioning etc. It allow to use desired application over the cloud. Application maintenance, security and updates is not consumer nuisance it is a responsibility of cloud service providers. Consumer only need browser to connect with it. There are many examples of SaaS vendors – Salesforce.com, Google Apps, Ning, Cenzic and etc.

Figure 2. Cloud Computing Services

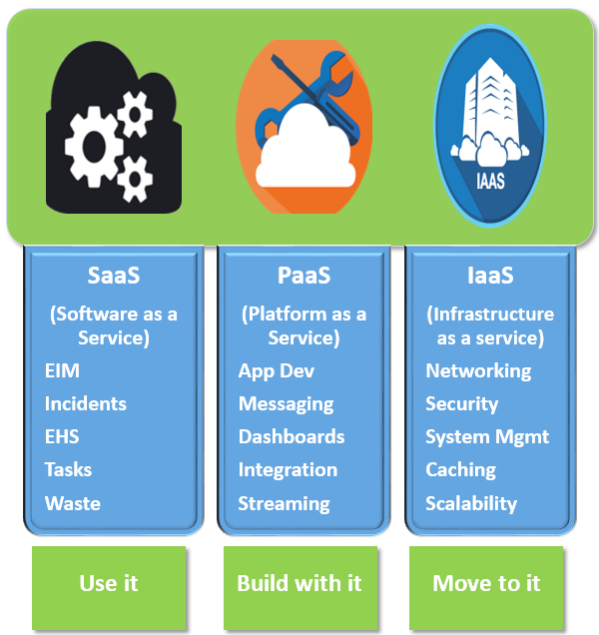

### C. Equations

CC services can be deployed in three different ways.
- Public cloud: In public cloud after deployment of your data you are not responsible to manage, maintain and host your data in the datacentre.

- Private Cloud: These clouds are also called as enterprise clouds, it is managed and maintain via internal sources in a private environment.

- Hybrid Cloud): It is a combination of public and private clouds. Data can be moved among both clouds with permission it provides flexible environment.

### III. IMPORTANCE OF CLOUD MIGRATION

In every organization database technology is fundamental tool. The need and importance of databases are growing with the growth of IT technology. Powerful database systems have been introduced. As the world is becoming a global village, many organizations are handling their business from remote locations. As this is an era of cloud computing organization keep their data in the cloud, moreover there is a need arise to migrate data among the clouds. Whereas Data migration is a process where we extract, clean, transform and upload data into the new platform [8].

### A. Cloud Migration Process:

These few steps will be followed by cloud migration process shown in Figure 3:

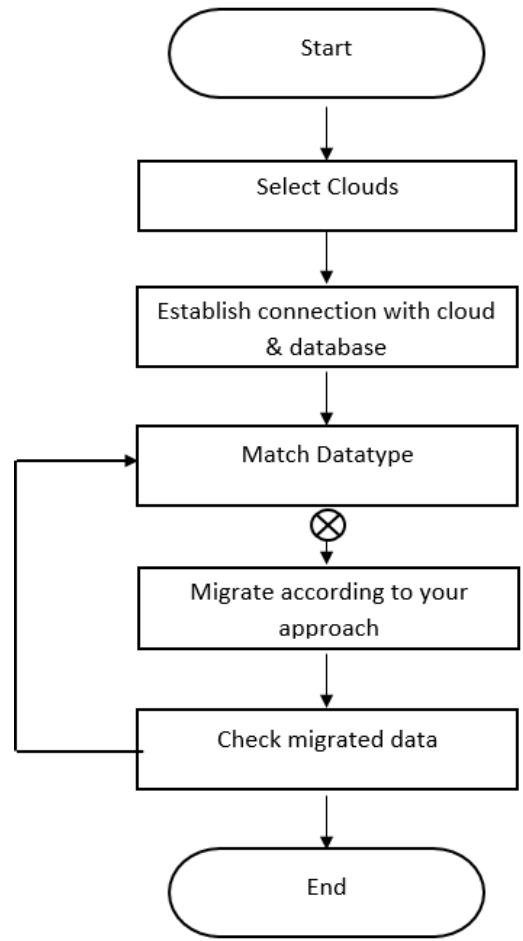

Figure 3. Cloud Migration Procedure [38]

Before migration below mention parameters of the database must be examined.
- Format Check: Check level of consistency and usability of the database.

- Consistent data: Exclude data which disrupt the logical consistency during the commit procedures.

- Length Check: examine length of your data when converting from one type to another.

- Range Check: Scrutinize the length of your data to assign an appropriate data type for memory efficiency.

- Integrity Check: Check integrities for the data association scheme.

## B. Data Migration Approaches

Previously, many approaches used for migration of data between systems. Many tools are fabricated with DBMS to accomplish this. Migration categorizes in three classes:
(a) By tools
(b) Manual data Migration
(c) User define new System for data migration and etc

According to research published in 2017 following technologies used by the developers cloud migration [7]. Mention in TABLE I.

TABLE I
TECHNOLOGIES USED FOR CLOUD MIGRATION

| 1 | Manual Coding | 174/**64%** |
|---|---|---|
| 2 | Data Integration Tool | 145/**54%** |
| 3 | Data Mapping Tool | 96/**36%** |
| 4 | Data Dictionary | 73/**27%** |
| 5 | Testing Tool | 52/**19%** |
| 6 | Data Archiving Tool | 29/**11%** |
| 7 | Other | 15/**16%** |

In our research, we want transfer data among clouds so we will use a third approach means we will program our user defined system to transfer data among the clouds.

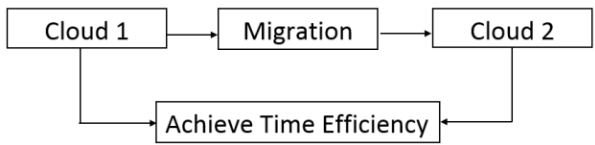

Figure 4. Cloud Migration Flow

## IV. PROBLEM STATEMENT

When there is data transfer between cloud rate of accuracy and speed is a main challenge. Previous researches transfer table data in columns and Obtained time efficiency [38]. In this research text files as well as image files would be transferred among cloud to accomplish time efficiency. Flow of migration strategy shown in Fig. 4.

### A. Data Integration:

Data integration is a scenario where data from various sources combine at one place, and produce a unified and meaningful view of data. Design such a system with maximum efficiency is a challenge. It is usually characterized by an architecture.
(a) Architecture of global schema
(b) Architecture of set of sources

Global schema contains a virtual view of data while the sources contain a the real data [42]. The difficult part in designing a data integration system is to design source depiction writing, mapping and schema which require expertise knowledge to write scripts [34].

Factors for Data Integration:
(a) XML
(b) Elastic Query Processing
(c) Model Organization
(d) Data organization
(e) P2P Data Organization

- XML (extensible mark-up language): XML is one of the pillars in the development of data integration. Reason behind this is its syntactic format so data can easily share among sources. For good data integration the system should capable of1) XML (extensible mark-up language): XML is one of the pillars in the development of data integration. Reason behind this is its syntactic format so data can easily share among sources. For good data integration the system should capable of handling complex XML. So it was challenging to design such a system which can interpret Nested and complex HTML tags [40]

- Elastic query processing: When a query posted on a schema and its task is to gather data from a set of sources to create a view. In order to form that view the query must be very efficient. There are many techniques available for efficient query processing, but in data integration system these techniques not completely applicable because its optimizer not contained detailed information so good path selection is not possible sometimes for query execution.

- Model Organization: For management of data integration system main task is to design a mapping route among schemas. So algebraic operation would be a solution to map between schemas, but it is a complex task.

- P2P Data Organization: Peer to peer is an architecture which shares data by means of distributed contrivance. In a distributed system complex semantic mappings are programmed individually among a set of sources and global schema network paths.

### B. Data Access Technique

These are data mapping techniques which used to access data from one or more sources. The data can be accessed from tables in a different manner. Here we will discuss about few data mapping techniques to access data
(a) Semantic Mapping
(b) Data Driven Mapping
(c) Object Related Mapping

- Semantic Mapping: This technique detects and discover exact matches. Any logical alteration of column data cannot be handled and recognized by it. Semantic system signifies information associated with a particular domain and concepts. It can provide, capable and potential domain dependent system schema to access data [10]. This is a process where an object, action, identities devoted to entities and algebraic sequences mapped them shown in Figure 5. The disadvantage of this technique is that it is domain dependent, does not support logic implementation of data and do not consult the metadata registry for synonym search.

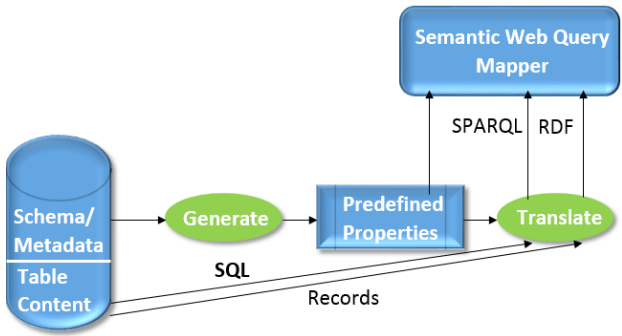

Figure 5. Semantic Mapping [43]

- Data-Driven Mapping: This technique evaluates data two times. (a) Heuristic evaluation (b) Statistical evaluation, so it is capable of automatic discovery complex mapping routes among a set of sources for data access and data integration [10]. With this approach set of sources can produce results which contain concatenations, arithmetic calculations, logical transformations, substrings, other data manipulation operations as shown in Figure 6. It can also define and handle exceptions.

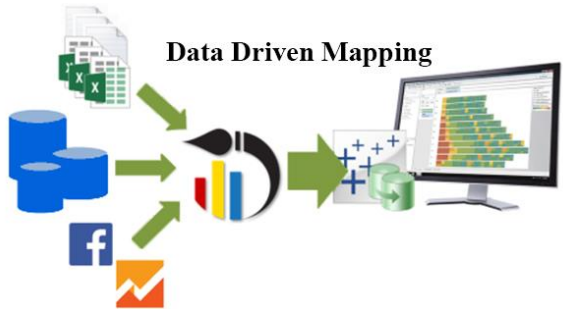

Figure 6. Data Driven Mapping [45]

- ORM Mapping: While need arises to transfer data between systems which are not compatible with each other, then the technique Object Relational Mapping (ORM) is used. Object oriented programming is used to create maps with relational database systems, XML repositories and other data sources, etc. Virtual object databases, create with this technique so data can be mapped and access [10]. Mostly database system deals with scalar values for extract, transform, load (ETL) and data manipulation operations so it is necessary that the object must behave both as an object and scalar value as per as need. As we have a logical representation of the object here so difficult part is that to store them in the database as an object and it should be capable of preserving their properties and its relationship with entities for reuse as scenario described in Figure 7(a) and Figure 7(b).

Figure 7(a). Object Related Mapping

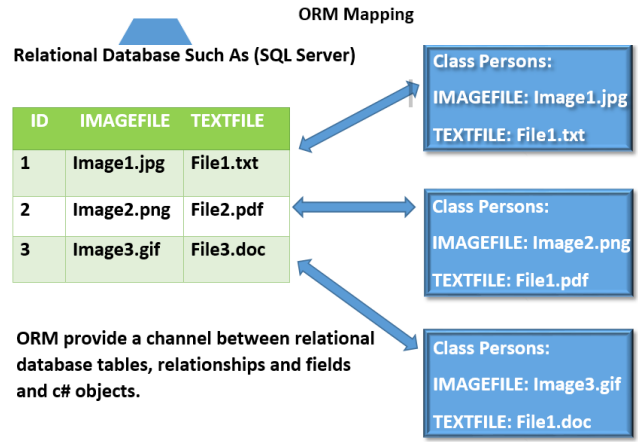

Figure 7(b). Object Related Mapping

## V. METHODOLOGY

For migration of data from azure cloud to gearhost cloud so we are using ORM technique. We have interact with object for efficient data migration. We will follow the entity data model and entity frame core (EF core) technology as shown in Fig. 9 for our research work.

### A. Entity Data Model

This model do not consider in which form data is stored it focuses on a concept regarding the structure of data.

- Conceptual Schema: This defines the scenario where we examine our entities, and decide which entity classes are required to accomplish tasks and establish relationship among them. By following these steps here we will produce a high level view of our database.

- Mapping: Both designed conceptual and storage schema will be mapped at this stage.

This model describes a relational model and data storage representation in Figure 8. Processing performed using Conceptual schema definition language (CSDL), mapping specification language (MSL), and store schema definition language (SSDL).

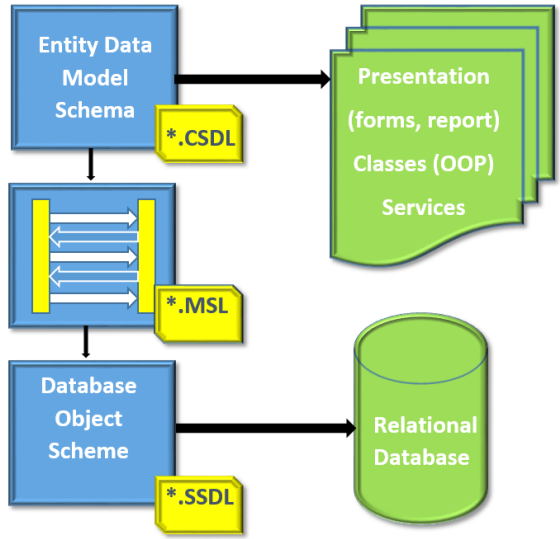

Figure 8 .Entity Relationship Diagram [44]

## A. EDM Data Structure Concepts

This model do not consider in which form data is stored it focuses on a concept regarding the structure of data.

- Entity: Structure of data describes in EDM come under the umbrella of this concept.
- Association: Referential concepts in tables describe through different association type in our database. Like Primary key, Foreign key, composite key, etc.
- Property type: Here we define property to the pre-defined entities. Like here we will define the data type of the columns, which can be primitive or complex.

## B. Data Loading

For loading data into application from your database we have three types of loadings.
(a)   Lazy Loading
(b)   Explicit Loading
(c)   Eager Loading.

- Lazy Loading: This is default data loading. It loads the main entity reference in the query instead of related entity reference in the query, which makes it slow.
- Explicit Loading: If we disable lazy loading so we can manually add related entity references to the query using load () method.
- Eager Loading: It loads all the related entities. But after the loading of the main entity of query. For this purpose, its use includes () and then include () methods. Advantage of eager loading is that it extracts massive data in one query processing.

## VI. EF CORE( ENTITY FRAMECORE WORK)

It lightweight, extensible technology and it support cross platform development. It can also provide functionality of ORM mapper by using objects created in .net application code. It supports many database engines like SQL Server, SQLite, Oracle, Microsoft Access files, PostgreSQL and etc. In our research we are dealing with SQL Server. [34]

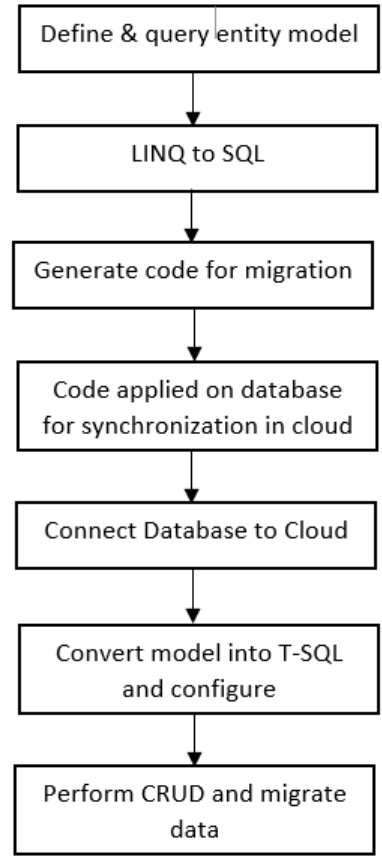

Figure 9 .Methodology

Steps Shown in Figure 9 are the operations we have perform to achieve cloud database migration.

## A. Prerequisite

Installed following application packages before starting configuration and ORM class and object defining scenario.
Microsoft.EntityFrameWorkcore.SqlServer
Microsoft.EntityFrameWorkcore.Tools -Pre
Microsoft.EntityFrameWorkcore.SqlServer.Domain [34].

## B. Creating Model

We have Fluent API configuration for creating the entity model. The Model builder API used to configure it. For defining properties data annotations have been used. Entity model has metadata of type blog.

## C. Keys Constraint

Primary keys and other unique identifiers are defined using the model builder API without effecting entity classes.

D. *Value Generation*

Value generated by the user in the database every time we populate database entities with data and that value is saved using savechanges () method.

E. *Required Entities*

In our framework 7 entities have been defined PersonalID, LastName, FirstName, Address, City, TextFile, Picture.Among them only required entity is PersonalID entity .Remaining entities left optional in order to test result in three different categories (a) column text data transfer (b) mage file migration (c) text file migration.

F. *Relational Modelling*

: For relational database modeling of discussed framework this package (Microsoft.EntityFrameworkCore.Relational package) is installed.

G. *Table Mapping*

The above mentioned package is used for table mapping DbSet<TEntity>. If table entity is not defined in dbset context, then class name will use for mapping [34].

H. *Column Mapping*

This mapping is performed to map column, that after query which column data should fetch and save in table of other database after migration.

I. *Data Mapping*

Datatype of columns are also mapped among databases as we are using the same database in different clouds so we will map datatype and their maximum length. Here we are using dbo schema in our databases.

## VII. MIGRATION

We have used Migration builder API for data transfer by using Migration Builder Operations () following up method approach.

## VIII. RESULTS

In our research, we have migrated data from one cloud to another both having different architectures. We are using EFcore technology for this purpose and ORM technique for object mapping. We will observe time efficiency in results when we compare it with and without EFcore and ORM.

TABLE II
FORMULAE

| Formulae For Result Calculations |
|---|
| Save Time Efficiency =Save Old-Save ORM |
| Transfer Time Efficiency =Transfer Old-Transfer ORM |
| Total Time Efficiency =Save Time Efficiency+Transfer Time Efficiency |

Whereas stopwatch() method is used to calculate time consumed by query execution in milliseconds.

TABLE III
IMAGE GB TIME EFFICIENCY ANALYSIS

| ISTE(GB) | ITTE(GB) | ITTTE(GB) |
|---|---|---|
| 6478 ms | 13048 ms | 19526 ms |
| -994 ms | 1557 ms | 563 ms |
| 5236 ms | -1768 ms | 3468 ms |
| 662 ms | 18145 ms | 18807 ms |
| -269 ms | 11171 ms | 3153 ms |
| -1329 ms | 8838 ms | -2435 ms |
| 4072 ms | -4589 ms | -517 ms |
| 295ms | -3949 ms | -3654 ms |
| 1111 ms | -1315 ms | -204 ms |
| 854ms | 896ms | 1750ms |

ISTE=Image Save Time Efficiency, ITTE= Image Transfer Time Efficiency, ITTTE= Image Total Transfer Time Efficiency, GB=gigabytes, ms=milliseconds

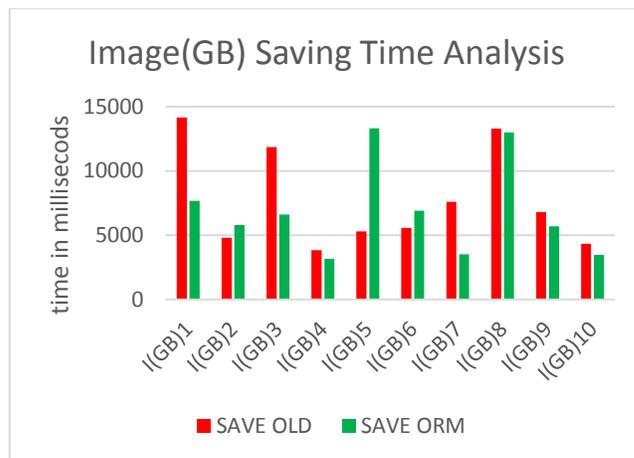

Figure 10. Image (GB) Saving Time Analysis, I(GB)=Image in GB

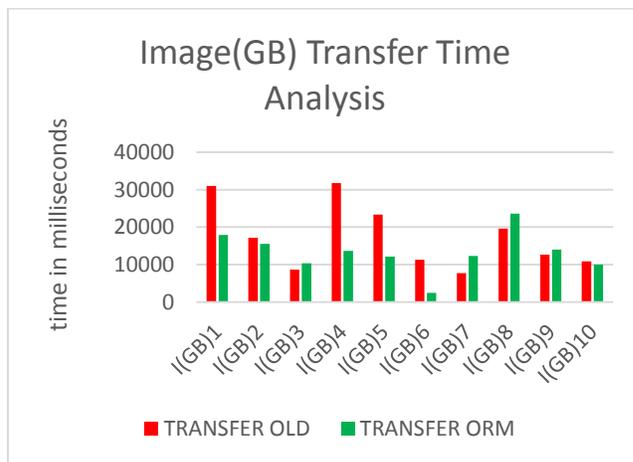

Figure 11. Image (GB) Transfer Time Analysis

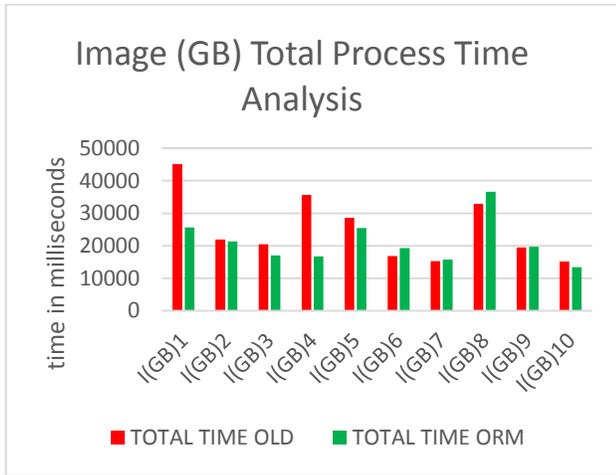

Figure 12. Image (GB) Total Process Time Analysis

TABLE IV
IMAGE KB TIME EFFICIENCY ANALYSIS

| ISTE(KB) | ITTE(KB) | ITTTE(KB) |
|---|---|---|
| -74 ms | -508 ms | -582 ms |
| 2516 ms | 1511 ms | 2326 ms |
| -2126 ms | 1114 ms | -1012 ms |
| 1364 ms | 8603 ms | 9967 ms |
| 375 ms | 120 ms | 495 ms |
| 577 ms | 3649 ms | 4226 ms |
| 1077 ms | 5921 ms | 6998 ms |
| 4735 ms | 2228 ms | 6963 ms |
| 1059 ms | 3159 ms | 4218 ms |
| 412 ms | 1212 ms | 1625 ms |

ISTE=Image Save Time Efficiency, ITTE= Image Transfer Time Efficiency, ITTTE= Image Total Transfer Time Efficiency, KB=kilobytes, ms=milliseconds

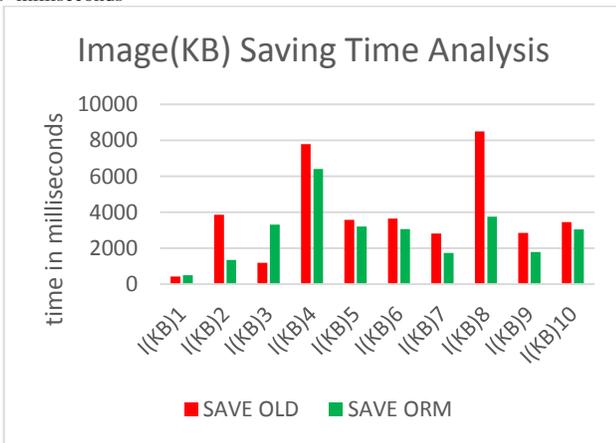

Figure 13. File (KB) Saving Time Analysis, I(kB)=Image in KB

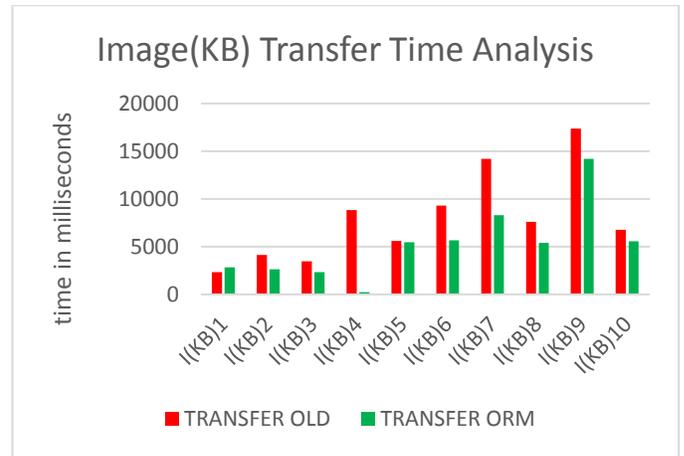

Figure 14. File (KB) Transfer Time Analysis

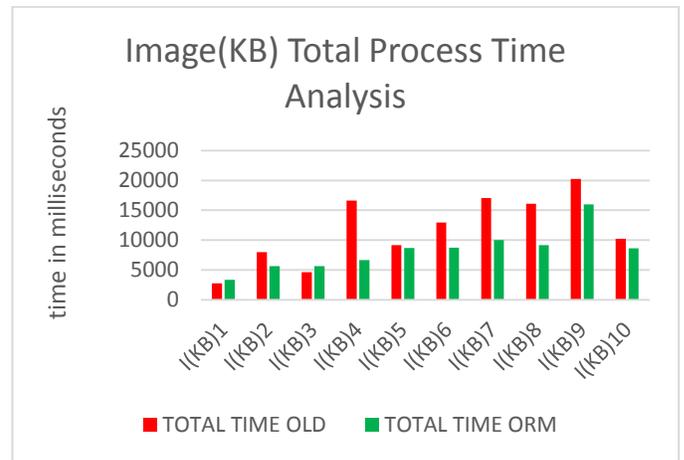

Figure 15. File (KB) Total Process Time Analysis

TABLE V
FILE KB TIME EFFICIENCY ANALYSIS

| FSTE(kb) | FTTE(kb) | FTTTE(kb) |
|---|---|---|
| -167ms | 427 ms | 260 ms |
| 936 ms | -516 ms | 420 ms |
| 333 ms | 829 ms | 1162 ms |
| 2871 ms | 975 ms | 3846 ms |
| 3309 ms | -560 ms | 2749 ms |
| 672 ms | -236 ms | 436 ms |
| 1816 ms | 222 ms | 2038 ms |
| -207 ms | -996 ms | -1203 ms |
| 1446 ms | -446 ms | 1000 ms |
| 81 ms | 140 ms | 221 ms |

FSTE=File Save Time Efficiency, FTTE= File Transfer Time Efficiency, FTTTE= File Total Transfer Time Efficiency, KB=kilobytes, ms=milliseconds

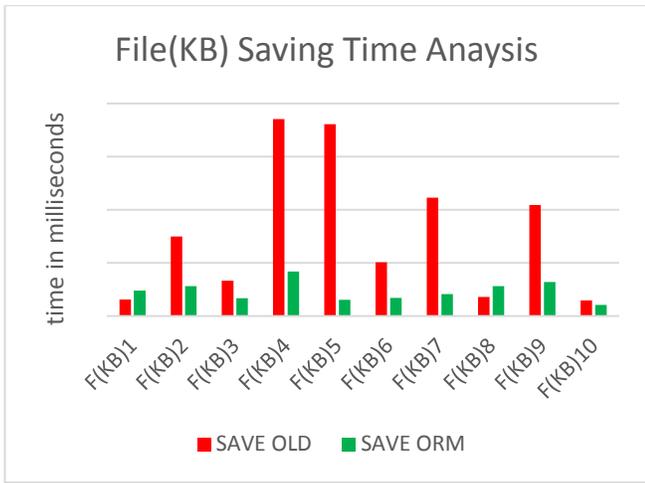

Figure 16. File (KB) Saving Time Analysis, F(MB)=File in KB

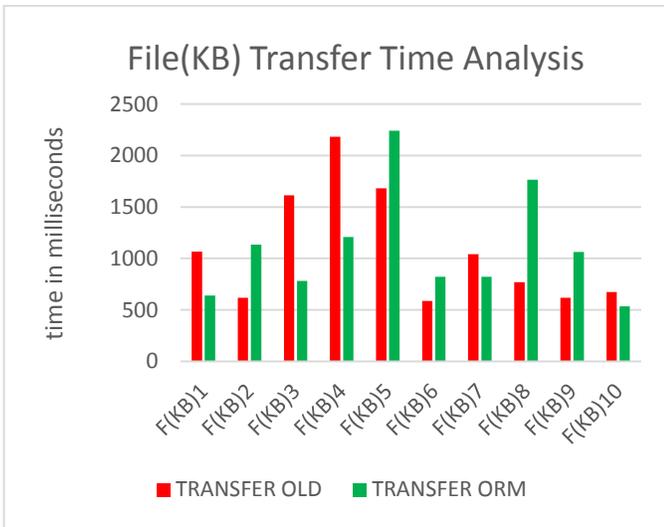

Figure 17. File (KB) Transfer Time Analysis

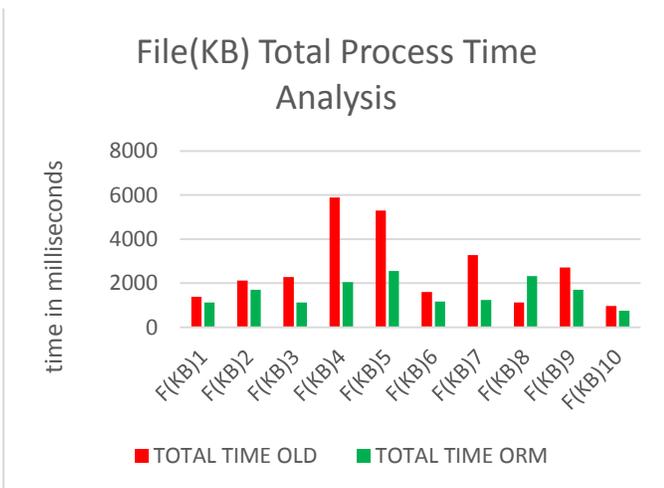

Figure 18. File (KB) Total Process Time Analysis

TABLE VI
FILE GB TIME EFFICIENCY ANALYSIS

| FSTE(GB) | FTTE(GB) | FTTTE(GB) |
|---|---|---|
| -1089 ms | 131 ms | 1473 ms |
| 23 ms | 1126 ms | 3142 ms |
| 9680 ms | 198 ms | 10696 ms |
| -33 ms | 121 ms | 1030 ms |
| 10439 ms | -608 ms | 11605 ms |
| 37 ms | 12 ms | 938 ms |
| 9901 ms | 223 ms | 10936 ms |
| 38 ms | 321 ms | 1260 ms |
| -78 ms | 420 ms | 1384 ms |
| 151 ms | -225 ms | 74 ms |

FSTE=File Save Time Efficiency, FTTE= File Transfer Time Efficiency, FTTTE= File Total Transfer Time Efficiency, GB=gigabytes, ms=milliseconds

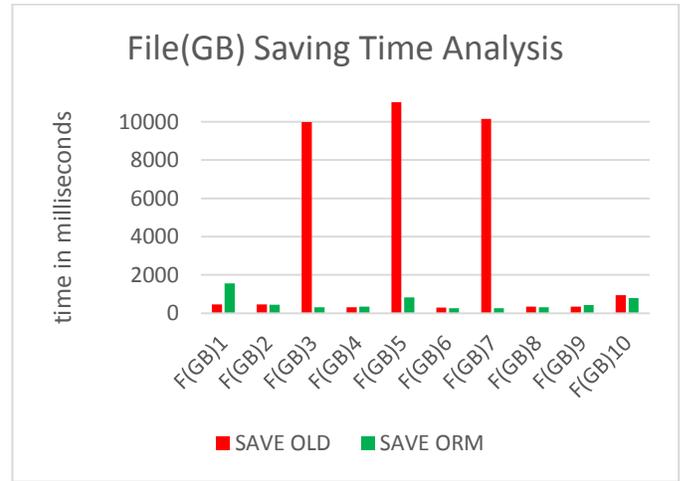

Figure 19. File (GB) Saving Time Analysis, (GB)=File in GB

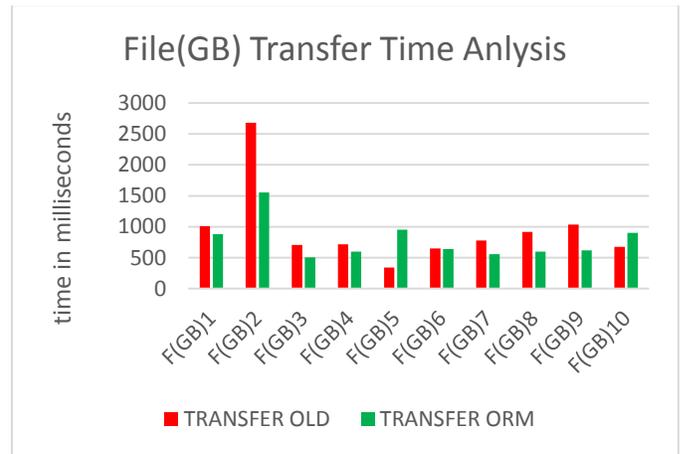

Figure 20. File (GB) Transfer Time Analysis

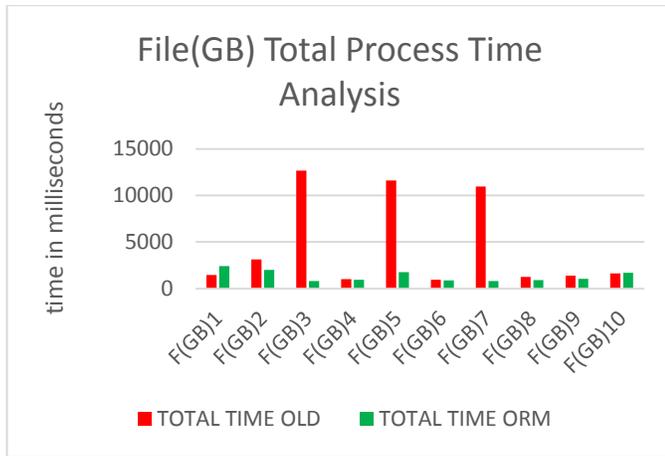

Figure 21. File (GB) Total Process Time Analysis

Results calculated in four categories (a) images in MB (b) images in KB (c) Text Files in MB (d) Text Files in KB. We have tested each category with 100 input files. But here we are showing results of only 10 individual cases in each category to demonstrate efficiency because it is not possible to show the results of all input files also taking the average of these input files can mislead to a conclusion.

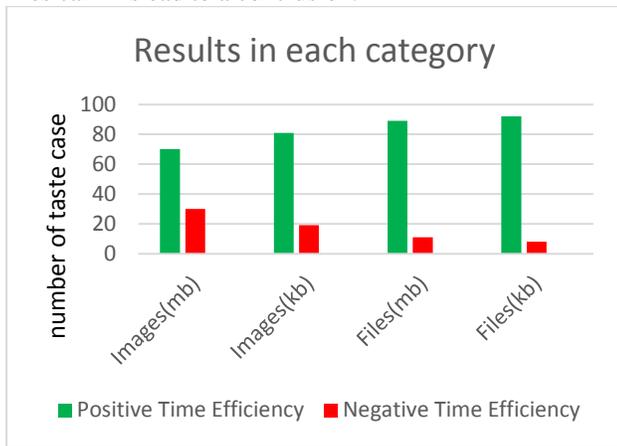

Figure 22. Aggregate Result Analysis

The above chart shows the number of successful and failed Cases in each category.

## IX. DISCUSSION

As we have mentioned values in tables from TABLE II to TABLE V and graphs can be observed from Fig. 10 to Fig. 21 in most of the cases ORM method is efficient then old method but in some cases very attention-grabbing results are found when difference in their time is more than 200% or more than that as we can see in Fig. 10. old method is taking very much less time in migration than ORM also investigating Fig. 19 found that old method is consuming very high time where as new method consumed few milliseconds for it. But interesting thing is that such cases are found when we are dealing with Images in GB and Images in KB.

## X. CONCLUSION

From the above results multiple things can be observed like sometimes Save Time Efficiency is negative and Sometimes Transfer Time Efficiency is negative so both of these factors are responsible for negative results. Whereas, one more important factor is speed of internet if good internet speed is constant, results can be more constant. Also image migration consuming more time than text file migration. If we combine the results of these 4 categories statistically we can say that we have achieved approximately 80% efficiency.

## XI. FUTURE WORK

It is still an ongoing research area in future instead of two clouds, multiple clouds can be used and test results, to find out the appropriate reason of negative efficiency. As we have transfer input from 100kb to 8GB. One can target to transfer images and files more than 8 GB to achieve positive time efficiency.

ACKNOWLEDGMENT

I would like to thank Muhammad Bilal Amjad from Microsoft for providing research resources of azure cloud.